\documentclass{elsart}

\def\crpar{\mbox{$\not\!\partial$}}

\begin{document}
\begin{frontmatter}


\title{Hopf term induced by fermions}
\author{Alexander G. Abanov}
\address{12-105, Department of Physics,
MIT, \\ 77 Massachusetts Ave.,
Cambridge, MA 02139, U.S.A.}


\begin{abstract}
    We derive an effective action for Dirac fermions coupled to O(3)
    non-linear $\sigma$-model (\mbox{NL$\sigma$M}) through the
    Yukawa-type interaction.  The nonperturbative (global) quantum
    anomaly of this model results in a Hopf term for the effective
    \mbox{NL$\sigma$M}.  We obtain this term using the ``embedding''
    of the CP$^1$ model into the CP$^{\rm M}$ generalization of the model
    which makes the quantum anomaly perturbative.  This perturbative
    anomaly is calculated by means of a gradient expansion of a
    fermionic determinant and is given by the Chern-Simons term for an
    auxiliary gauge field.
\end{abstract}
\begin{keyword}
quantum anomalies, fermionic determinant, Hopf term, effective action
\end{keyword}
\end{frontmatter}

It is well-known that
nonperturbative anomalies\cite{Witten-1982SU(2)} in gauge theories can
be reduced to perturbative ones by embedding the gauge group into a
bigger one\cite{ElitzurNair-1984,Klinkhamer-1991}. In this paper we
show that a similar method can be used to calculate the effect of
nonperturbative anomalies on the effective action of a non-linear
$\sigma$ model induced by fermions.

We consider an effective action $S_{eff}(n)$ of Dirac fermions on the 
three-dimensional sphere $S^3$ coupled to a background chiral field $n$:
\begin{equation}
    e^{-S_{eff}(n)} = \int d\psi\,d\bar\psi\,
    \exp\left(-\int_{S^{3}}d^{3}x\, \, \bar{\psi}\left[i \crpar 
     + im \hat{n}\right]\psi\right).
 \label{eq:model}
\end{equation}
Here  $\crpar = \gamma^{\mu}\partial_{\mu}$ where the $\gamma^{\mu}$ are
three-dimensional gamma-matrices which can be chosen, e.g., to be Pauli
matrices, $\hat{n}=\vec{n}\cdot\vec{\tau}$ with $\vec{n}\in S^2$,
$\vec{n}^2=1$,  $\vec{\tau}$ is a set of Pauli matrices acting in
the isospace, and we use a Euclidian formulation.

We calculate the variation of the effective action $S_{eff}(n)=-\ln\det D$,
$D=i\crpar + im \hat{n}$  with respect to $n$.
\begin{equation}
    \delta S_{eff} = -\mbox{Tr}\, 
    \left[\delta D D^{\dagger} 
    (DD^{\dagger})^{-1}\right],
 \label{eq:var}
\end{equation}
where $\delta D=im\delta\hat{n}$ and $D^\dagger =i\crpar - im \hat{n}$.
Then we use\cite{AbanovWiegmann-2000} $DD^{\dagger} =
-\partial^{2}+m^{2}+m\crpar\hat{n}$ and expand in $m\crpar\hat{n}$
obtaining $(DD^{\dagger})^{-1} =
G_{0}-G_{0}m\crpar\hat{n}G_{0} + G_{0} (-m\crpar\hat{n}G_{0})^{2}+
\ldots$, where $G_{0}= \frac{1}{-\partial^{2}+m^{2}}$.

Calculating traces and leaving only first nonzero orders in $1/m$ of 
real and imaginary parts of the effective action we obtain (only trace
in isospace is left)
\begin{eqnarray}
    \delta S_{eff} & = & \delta S_{Re} + \delta S_{Im},
 \label{eq:duvar}  \\
    \delta S_{Re} & = & \frac{|m|}{8\pi}\,\int d^{3}x\, 
    \mbox{tr}\,(\partial_{\mu}\delta\hat{n}) 
    (\partial_{\mu}\hat{n}) +\ldots,
 \label{eq:effRe}  \\
    \delta S_{Im} & = & -i\frac{\mbox{sgn}(m)}{32\pi}\,\int d^{3}x\, 
    \epsilon^{\mu\nu\lambda}\,\mbox{tr}\, (\hat{n}\delta\hat{n} 
    \partial_{\mu}\hat{n} \partial_{\nu}\hat{n} 
    \partial_{\lambda}\hat{n}) .
 \label{eq:effIm}
\end{eqnarray}

Now our goal is to restore the effective action  from its variation
(\ref{eq:duvar}-\ref{eq:effIm}). For the real 
part we have
\begin{equation}
    S_{Re}  =  \frac{|m|}{16\pi}\,\int d^{3}x\, 
    \mbox{tr}\,(\partial_{\mu}\hat{n})^{2} +\ldots .
 \label{eq:effReFull}  
\end{equation}

It is straightforward to check that the variation of the imaginary
part of the action (\ref{eq:effIm}) is identically zero and naively
$S_{Im}=0$. However, because of the nontrivial homotopy group
$\pi_3(S^2)=Z$, the configurations of $n$ are divided into topological
classes\cite{DFN-1985} labeled by an integer-valued Hopf invariant
$H(n)$. If $S_{Im}\sim H(n)$ then $\delta S_{Im}=0$ and we can not
find the ``Hopf term'' from our perturbative calculation. The presence
of the Hopf term in the effective action is of great importance
because it changes spin and statistics of solitons of
\mbox{NL$\sigma$M}\cite{WilczekZee-1983}.

In the following we show that the correct result for the imaginary part is
\begin{equation}
    S_{Im}  =  -i\pi\, \mbox{sgn}(m)\, H(n).
 \label{eq:effImFull}  
\end{equation}
The value of the coefficient in front of the Hopf invariant
corresponds to the Fermi-Dirac statistics of solitons which agrees
with their fermionic charge\cite{Jar-1987,AbanovWiegmann-2000}.

To find the imaginary part of the effective action we generalize the
model (\ref{eq:model}), changing the size of isospace and replacing
$\hat{n}$ by $\hat{n}=2zz^\dagger -1$ with
$z^{t}=(z_{1},z_{2},\ldots,z_{M+1})$ -- complex vector with unit
modulus $z^{\dagger}z=1$. The $(M+1)\times (M+1)$ matrix $\hat{n}$
does not depend on the phase of $z$, i.e., $z$ should be considered as
a CP$^{\rm M}$ field\cite{CPM}. We note that for the particular
configuration $z^{t}=(z_{1},z_{2},0,\ldots,0)$ the fermions with
isospace indices higher than $2$ are decoupled from $z$ and do not
contribute $z$-dependent terms into the effective action.  As a
consequence we can obtain the effective action for (\ref{eq:model}) by
restricting the effective action of the CP$^{\rm M}$ model to particular
configurations $z^{t}=(z_{1},z_{2},0,\ldots,0)$.

It is easy to check that $\hat{n}^{2}=(2zz^\dagger-1)^2=1$ and
perturbative results (\ref{eq:duvar}-\ref{eq:effIm}) are still
valid. However, for $M>1$ the homotopy 
group $\pi_3(\mbox{CP}^{\rm M})=0$ and there
are no topologically nontrivial configurations of $\hat{n}$. The
effective action can be found perturbatively!  Substituting
$\hat{n}=2zz^{\dagger}-1$ into (\ref{eq:effIm}) we obtain after some
algebra
\begin{equation}
    \delta S_{Im} = i\frac{\mbox{sgn}(m)}{2\pi}\,\int d^{3}x\,
    \epsilon^{\mu\nu\lambda}\,\delta a_{\mu}\,\partial_{\nu}a_{\lambda},
 \label{eq:SImvar}
\end{equation}
where $a_{\mu}=z^{\dagger}(-i\partial_{\mu})z$.
From here we obtain
\begin{equation}
    S_{Im} = i\frac{\mbox{sgn}(m)}{4\pi}\,\int d^{3}x\,
    \epsilon^{\mu\nu\lambda}\, a_{\mu}\,\partial_{\nu}a_{\lambda}.
 \label{eq:effImFullpert}  
\end{equation}

Restricting (\ref{eq:effImFullpert}) to particular CP$^1=S^2$
configurations we have (\ref{eq:effImFull}) with the well-known
expression for the Hopf invariant\cite{DFN-1985}
\begin{equation}
    H(n) = -\frac{1}{4\pi^2}\,\int d^{3}x\,
    \epsilon^{\mu\nu\lambda}\, a_{\mu}\,\partial_{\nu}a_{\lambda},
  \label{eq:Hopf}  
\end{equation}
where $a_\mu=z^\dagger(-i\partial_\mu)z$ with a two-component $z$ and
$\vec{n}=z^\dagger\vec{\tau}z$.

Combining (\ref{eq:effReFull}) and (\ref{eq:effImFull}) we obtain for
the effective action of the CP$^{\rm M}$ non-linear $\sigma$-model induced
by Dirac fermions
\begin{equation}
    S_{eff} = \int d^{3}x\, \left\{ \frac{|m|}{16\pi}\,
    \mbox{tr}\,(\partial_{\mu}\hat{n})^{2}+i\frac{\mbox{sgn}(m)}{4\pi}
    \,  \epsilon^{\mu\nu\lambda}\, 
    a_{\mu}\,\partial_{\nu}a_{\lambda}\right\},
  \label{eq:SeffCPMFull}
\end{equation}
where we kept only the imaginary part of the action and the terms of
order $m$ in the real part. The second term of (\ref{eq:SeffCPMFull})
is a perturbative ``Chern-Simons'' term in the case of $M>1$. It
becomes a nonperturbative (global) Hopf term in the case of $M=1$.

In conclusion, we have derived  the effective
action for the CP$^{\rm M}$ non-linear $\sigma$-model induced by Dirac
fermions on a three-dimensional sphere.
We have shown that this effective action has a nontrivial topological
term which (for $M=1$) is equal to the well-known Hopf term for
the O(3) non-linear $\sigma$-model. We used the method of
embedding\cite{ElitzurNair-1984,Klinkhamer-1991} known for global
anomalies in gauge theories. 

We would like to point out some differences between
global terms for the $\sigma$-model and for the gauge field models.
In the case of gauge fields there always exists a direct
interpolation between  configurations which  differ by
non-trivial gauge transformation\cite{Witten-1982SU(2)}. Therefore,
the question of the relative phase of fermionic determinants for those 
configurations is well-defined. In the case of the \mbox{NL$\sigma$M} there is
no such direct interpolation between configurations from different
topological classes.
Therefore, strictly speaking, these topological classes can be weighed in 
the partition function corresponding to (\ref{eq:model}) with arbitrary 
weights. 
The imaginary part of an effective action, or the
relative phase of determinants with chiral field configurations 
belonging to different topological classes, is not defined. However, in 
a realistic physical model there might be some regularization which 
allows one to connect field configurations from different topological 
classes. E.g., if a theory is defined on a lattice, the singular 
processes changing topological classes 
are allowed. Another possibility is to make a constraint, $\hat{n}^{2}=1$, 
soft. Then at some points in space-time $\hat{n}=0$, the target 
manifold is not $S^{2}$, and there are no distinct 
topological classes anymore. 

In this sense the ``embedding method'' we used is not just a
technical trick but is essentially a
method of regularization which allows us to interpolate between different
topological classes of \mbox{NL$\sigma$M}. 

We thank M. Braverman for discussing  mathematics. We would also like
to thank  P.A. Lee,
X.-G. Wen, and especially  P.B. Wiegmann for many helpful discussions. 
We appreciate the  hospitality of Aspen Center for Physics where part
of this work has been done.
This research has been supported by NSF DMR 9813764.


\begin{thebibliography}{99}


\bibitem{Witten-1982SU(2)}
E. Witten, Phys. Lett. {\bf 117B} (1982) 324-328 \\
{\it An SU(2) Anomaly}

\bibitem{ElitzurNair-1984}
S. Elitzur and V.P. Nair, Nucl. Phys. {\bf B243} (1984) 205-211 \\
{\it Nonperturbative Anomalies in Higher Dimensions}

\bibitem{Klinkhamer-1991}
F.R. Klinkhamer, Phys. Lett. {\bf 256B} (1991) 41-42 \\
{\it Another look at the SU(2) anomaly}

\bibitem{AbanovWiegmann-2000}
A.G. Abanov and P.B. Wiegmann, Nucl. Phys. {\bf B570} (2000) 685-698 \\
{\it Theta-terms in non-linear sigma-models.}

\bibitem{DFN-1985}
B.A. Dubrovin, A.T. Fomenko, S.P. Novikov,
\\ {\it Modern Geometry-Methods and
Applications : Part II, the Geometry and Topology of Manifolds}
 (Graduate Texts in Mathematics, Vol 104), Springer-Verlag, 1985.

\bibitem{WilczekZee-1983}
F. Wilczek and A. Zee, Phys. Rev. Lett. {\bf 51} (1983) 2250-2252 
 \\ {\it Linking Numbers, Spin, and Statistics of Solitons}

\bibitem{Jar-1987}
T. Jaroszewicz, Phys. Lett.  {\bf B193} (1987) 479-485 \\
{\it Fermion-Induced Spin of Solitons:
                          Vacuum and Collective Aspects}

\bibitem{CPM}
The idea of using CP$^{\rm M}$ representation to derive Hopf term belongs to
T. Jaroszewicz\cite{Jar-1985}. However, he used ``gauge rotation''
with nonzero Jacobian and the derivation of Ref.\cite{Jar-1985} is
incorrect.

\bibitem{Jar-1985}
T. Jaroszewicz, Phys. Lett. {\bf B159} (1985) 299-302\\
{\it Induced Topological Terms,
	Spin and Statistics in (2+1) Dimensions}



\end{thebibliography}
\end{document}